\documentstyle[preprint,aps]{revtex}
\begin{document}
\draft
\title{Relative spins and excitation energies of
superdeformed
bands in $^{190}$Hg: Further evidence for
octupole vibration}

\author{B. Crowell,$^1$ M.P. Carpenter,$^1$ R.V.F. Janssens,$^1$
D.J. Blumenthal,$^1$ J. Timar,$^2$ A.N. Wilson,$^2$
J.F. Sharpey-Schafer,$^{3,2}$ T. Nakatsukasa,$^4$ I.
Ahmad,$^1$ A. Astier,$^5$  F. Azaiez,$^6$ L. du Croux,$^5$
 B.J.P. Gall,$^3$ F. Hannachi,$^7$ T.L. Khoo,$^1$ A.
Korichi,$^6$ T. Lauritsen,$^1$ A. Lopez-Martens,$^7$
M. Meyer,$^5$ D. Nisius,$^1$$^*$ E.S. Paul,$^2$ M.G. Porquet$^7$
and N. Redon$^5$}

\address{$^1$Argonne National Laboratory, Argonne, IL
60439;
$^2$University of Liverpool, Liverpool L69 3BX,
United Kingdom;
$^3$Centre de Recherches Nucleaires, F-67037
Strasbourg Cedex, France;
$^4$AECL Research, Chalk River Laboratories,
Chalk River, Ontario K0J 1J0, Canada;
$^5$Institut de Physique Nucl\'{e}aire Lyon,
IN2P3-CNRS, Lyon, F-69622 Villeurbanne, France;
$^6$I.P.N., IN2P3-CNRS b\^{a}t. 104-108, F-91405
Orsay Cedex, France;
$^7$C.S.N.S.M., IN2P3-CNRS b\^{a}t. 104, F-91406
Orsay Cedex, France}

\maketitle

\vspace{0.01in}

\begin{abstract}

An experiment using the Eurogam Phase II gamma-ray spectrometer
confirms the existence of an excited superdeformed
(SD) band in $^{190}$Hg and its very
unusual decay into the lowest SD band over
3-4 transitions.  The energies and dipole
character of the transitions linking the two
SD bands have been firmly established.  Comparisons
with RPA calculations indicate that the excited
SD band can be interpreted as an octupole-vibrational
structure.


\end{abstract}
\vspace{0.01in}
\thispagestyle{empty}

\vspace{0.01in}

\noindent
--------------- \\
\noindent
* graduate student from Purdue Univ., West Lafayette, IN 47907

\newpage

Recently, we have reported \cite{Cro94a} the
observation of an excited
superdeformed (SD) band in $^{190}$Hg with the
unusual property that it
decays entirely to the lowest SD band \cite{Dri90,Bea92}.
The only other cases
in even-even nuclei where transitions
linking two SD bands (i.e.
interband transitions) have been proposed are in
$^{150}$Gd and $^{152}$Dy
\cite{Fal94,Dag94}.  In all three nuclei,
the presence of the interband transitions
was inferred from the observed coincidence relationships.
In the case of $^{190}$Hg, the
excited SD band  was populated very weakly:
the gamma-ray intensities
ranged from 0.02 to 0.06\% with
respect to population of the ground state.
Hence, the possible interband
transitions given in Ref. \cite{Cro94a}
were reported as tentative.

In general, the direct measurement of
transitions between SD bands is
of great interest because it can give
critical information on the
properties of these nuclei at extreme
deformations. First, the fact
that such transitions are so uncommon
(excluding transitions between
signature partner bands, e.g.
Ref. \cite{Hg193}) suggests that
atypical intrinsic SD structures
are involved.  Second, the accurate
determination of relative energies,
of relative spins and parities, and
of branching ratios between
in-band  and out-of-band decay becomes
possible. This allows for
stringent new tests of mean-field
calculations whose results have thus
far mostly been compared  with measured
dynamic moments of
inertia, $\Im$$^{(2)}$.

	The SD minima in both the A$\sim$150 and
A$\sim$190 regions of superdeformation are
calculated \cite{SDOctupoles,Bon94}
to be soft with respect to octupole deformation.
Thus, low-lying octupole-vibrational states
should exist, and strong E1 transitions
are expected to connect these vibrational
levels to the lowest SD band.  The nuclei
in which these effects are likely to be observed
are those in which the excitation energy of
an octupole-vibrational band is low at the
spins at which SD bands are fed.  Calculations
using the random phase approximation
(RPA), described below,
predict that at a rotational frequency
$\hbar$$\omega$=0.4
MeV the lowest octupole-vibrational states in $^{190}$Hg,
$^{192}$Hg and $^{194}$Hg occur at 0.37, 0.67
and 0.67 MeV above the lowest SD band, respectively.
Thus, among the even-even Hg isotopes in
which superdeformation has been observed,
$^{190}$Hg is the best candidate for the observation
of octupole vibrations.  In Ref.
\cite{Cro94a} the excited band in $^{190}$Hg was
interpreted as an octupole vibration.
This interpretation is strengthened
here by the firm identification of the interband
transitions and the experimental determination
of their dipole character.

	The results presented here are from a new measurement
performed with the Eurogam Phase II gamma-ray spectrometer
\cite{Nol90}, which consists of 54 escape-suppressed
Ge detectors. Of these, 30 are large-volume
Ge detectors located in rings at 22, 46, 134
and 158 degrees with respect to the beam.
The remaining 24 detectors, at 75 and 105 degrees, are
so-called ``clover'' detectors, each consisting of four
closely packed Ge crystals within a single cryostat
inserted in an escape-suppression shield.  The
higher granularity of the ``clover'' detectors
reduces Doppler broadening
and allows measurements of linear polarization  \cite{Jon95}.
The experiment was performed using the reaction
$^{160}$Gd($^{34}$S,4n) on a stack of two
isotopically pure 0.5 mg/cm$^{2}$ Gd targets.
The newly commissioned Vivitron accelerator
was set to a beam energy of 153 MeV, but had
not been calibrated, and the actual energy
was estimated to be $\sim$158 MeV based on
a comparison of the measured gamma-ray spectra
with those from the earlier experiment performed
at the 88-inch cyclotron at Lawrence Berkeley
Laboratory \cite{Cro94a}.  An event was written
to tape if four or more escape-suppressed
Ge detectors recorded gamma rays in coincidence.
A total of 4.4$\times$10$^{8}$ events was acquired.
When two crystals of a ``clover'' detector
both registered signals,  their energies were
added in software, and the detector was treated
as a single unit.  The Doppler correction in such
cases was calculated under the assumption that the
primary gamma-ray entered the crystal that recorded
the higher energy.  When more than two crystals
of a ``clover'' detector recorded signals, the
information from the entire
detector was disregarded for that event; cases such
as these were found to arise mostly from pileup.
Unfolding the coincidence events into all
possible combinations of 3 and 4 energies
yielded, respectively, 2.5$\times$10$^9$ triples
($\gamma$$^{3}$) and 2.0$\times$10$^9$ quadruples
($\gamma$$^{4}$) combinations.  The data were
subsequently analyzed by sorting all the $\gamma$$^{3}$
combinations into a three-dimensional histogram
\cite{Kue92} and then extracting double-gated
one-dimensional spectra with full background-subtraction
and propagation of errors \cite{Cro94b}.

	A coincidence spectrum showing the excited
SD band and its decay into the lowest  SD
band is presented in Fig. 1.  The quality
of the spectrum is greatly improved with respect
to those shown in Ref. \cite{Cro94a}, both
because of the factor of four increase in
statistics and because of the use of a lower
beam energy, which avoided a number of interfering
gamma-ray transitions from $^{188}$Hg.  The
sensitivity to dipole transitions was also
enhanced by the presence of more detectors
at angles close to 90$^{\circ}$, where the
angular distribution is peaked for this type
of transition.  As can be seen from Fig. 1,
this experiment confirms the existence of
the excited SD band, as well as its
unusual pattern of decay to the lowest SD
band.  The transitions stand out clearly
above the background, and the corresponding
energies were determined more accurately
(Fig. 2).  The weak transitions at the top
and bottom of the band are also seen in the
present data with significantly better ratios
of signal to noise.

More importantly, the
interband transitions are now firmly established.
 The gamma-ray energies tentatively assigned
in Ref. \cite{Cro94a} to the interband transitions
were found to be incorrect, and the actual
transitions have energies of 812, 864 and
911 keV.  The relative positions of the two
SD bands is presented in Fig. 2.  Compared
with Ref. \cite{Cro94a}, the levels of the
excited SD band have been shifted up by one
transition, i.e. the excited band decays into
the lowest SD band one level higher than previously
thought.  Based on the present data,
the relative placement of the two bands
is uniquely determined by the sums and differences
of gamma-ray energies and by coincidence measurements.
For example, the 911 keV transition is observed
in coincidence with all members of the excited band
down to the 511 keV transition, and with all members
of the lowest SD band up to the 521 keV line, but not
with the higher members.
Weak evidence was also found for
an additional linking transition, with an
energy of 757 keV, lying above the three others
(dashed arrow in Fig. 2), but its intensity
was at the limit of the sensitivity of the
experiment, leading only to an upper limit
for this branch.  The energies of the transitions
in the lowest SD band were also measured more accurately
than in Ref. \cite{Cro94a}.  These energies are:
316.9(4), 360 (unresolved doublet), 402.34(4),
442.98(6), 482.71(6), 521.30(6), 558.6(1),
594.9(1), 630.1(1), 664.1(1), 696.9(1), 728.5(4)
757.4(4), 783.5(6),
and 801.8(8) (tentative).

	In order to determine the relative spins
of the two SD bands, DCO ratios (Directional
Correlations from Oriented nuclei \cite{Kra73})
were extracted for the interband transitions.
 These were defined simply as the number of
counts in the clover detectors (at side
angles) divided by the number of counts in
the large-volume detectors (at forward and
backward angles), after efficiency correction.
 These ratios were calibrated using transitions
of known multipolarity, and were found to
be approximately 1.3 for E1 transitions
and 0.8 for E2 transitions.  The ratios
were extracted from single-gated and double-gated
spectra under the assumption that the requirement
of a gating transition or transitions did
not significantly perturb the orientation
of the nuclear angular momentum.  This is
a valid assumption because the gating transitions
were measured at all angles by the nearly
spherically symmetric array of detectors,
and because the perturbation is negligible
for $\lambda$$\ll$J, where $\lambda$ is
the multipolarity of
the gating transition \cite{Bar87}.  The measured
DCO ratios for the interband transitions
are shown in Fig. 3, along with representative
E1 and E2 transitions.  Although the error bars are large,
the clustering of these values around those
of other dipole transitions indicates that
they are of dipole character.
Thus, it is highly probable that
the two SD bands differ by one unit of
angular momentum.

	A measurement of the polarizations of the
interband transitions was also attempted using
the clover detectors, based on the asymmetry
between Compton scattering parallel and perpendicular
to the reaction plane.  Although the technique
was successful for transitions with strengths
as low as a few percent relative to the $^{190}$Hg
channel, the statistics available in this
experiment were not sufficient to allow such
a measurement for the very weak interband
transitions of interest.

	Assuming that the excited band has a transition
quadrupole moment of 18(3) eb, equal to that
of the lowest SD band \cite{Dri90}, it is possible
to extract the partial half-lives of the interband
transitions. These are shown in Table I, along
with the transition strengths in units of
Weisskopf units (W.u.) under the assumptions of
E1 and M1 multipolarity.  As argued
in Ref. \cite{Cro94a}, it is unlikely that
M1 transitions between different quasiparticle
configurations would occur with such short
partial half-lives; the B(M1) values in Table I
are much larger than those typically observed
in deformed nuclei \cite{Lob68}.  Therefore,
these transitions
are very likely of E1 character.
The B(E1) values in Table I are about
two orders of magnitude stronger than
typical E1 transitions in heavy nuclei \cite{Lob68},
but are similar to those observed among
actinide nuclei exhibiting strong octupole
collectivity in the normally deformed well
\cite{Ahm93}.

	These findings reinforce the arguments put
forward in Ref. \cite{Cro94a}, i.e. there
is strong evidence that these states are of
octupole-vibrational character (for the range
of rotational frequencies over which E1 transitions
are observed), rather than members of a negative-parity
two-quasiparticle band.  We have therefore
carried out RPA calculations based on the cranked Nilsson model
for the octupole-vibrational modes.
The equilibrium quadrupole deformation,
$\delta_{\rm osc}=0.454$, and
pairing gaps,
$(\Delta_n,\Delta_p)=(0.8,0.6)$ MeV,
were determined by means
of the conventional shell correction method
with smoothed pairing gaps \cite{Bra72}.  The
form of the Hamiltonian is the same as in
Ref. \cite{Nak94}. A more detailed discussion
of the calculations will be given in a future
paper \cite{Nak95}.  All calculated quantities
given here for the excited SD band, including
Routhians and aligned angular momenta, are differences
between the relevant value for the excited band
itself and that of the lowest SD band.

	A comparison between the observed and calculated
$\Im$$^{(2)}$ moments of inertia for the excited
band is shown in Fig. 4a.  The excited SD
band has an average $\Im$$^{(2)}$ value of
123 $\hbar$$^{2}$MeV$^{-1}$, which is unusually
large for an SD band in an even-even A$\sim$190
nucleus. It is essentially constant over the
entire range of frequencies where the band
is observed.  The calculations not only accurately
reproduce the measured $\Im$$^{(2)}$, but
also match well with the observed excitation
energy of the vibrational band relative to
the lowest band (Fig. 4b).  (It should be
noted that although generator coordinate calculations
have not been carried out for $^{190}$Hg,
such calculations\cite{Bon94} predict phonon
energies for nearby nuclei which are nearly
twice as large as those resulting from the
RPA calculations.)  Finally, preliminary calculations
of the E1 transition rates yield values of
the order of 10$^{-3}$ to 10$^{-4}$ Weisskopf units,
consistent with the observed rates.

	The agreement found above between theory
and experiment makes it interesting to examine
the calculated wave-functions for insight
into the physical structure of the observed
states in the excited SD band.  The structure
of the lowest excited RPA state evolves smoothly
as a function of rotational frequency, but
for purposes of discussion it is convenient
to define three regions of frequency:
$\hbar$$\omega$=0 -- 0.2 MeV,  0.2 --
0.4 MeV, and above 0.4 MeV.  In the first
region, the RPA states make up an octupole-vibrational
multiplet.  At zero frequency, the K=2, 0,
3 and 1 modes occur at 0.99, 1.28, 1.35 and
1.41 MeV, respectively.  These modes become
increasingly mixed by the Coriolis force as
the rotational frequency increases.  Thus, in the
second region of frequency, which corresponds
roughly to that at which the SD band is observed
experimentally, the lowest vibrational state
becomes rotationally aligned.
At $\hbar$$\omega$=0.35 MeV,
its aligned
angular momentum (relative to that of the
lowest SD band) reaches a maximum value
of $\sim$3$\hbar$,  which corresponds
to a completely rotationally aligned octupole
phonon.  The similarity between the slopes
of the theoretical and experimental Routhians in
Fig. 4b indicates that the calculations reproduce the
observed alignment of the excited band.
In the highest region of frequency,
the lowest RPA state gradually loses its collective
vibrational character, eventually becoming
a two-quasineutron state.  The low excitation
energy of the octupole-vibrational band at
high spins in $^{190}$Hg compared to $^{192}$Hg
and $^{194}$Hg is attributable to the much
lower phonon energies of the low-K members
of the octupole-vibrational multiplet in this
nucleus.  More precisely, the close spacing in
energy of the different modes leads to stronger
Coriolis mixing and a smaller excitation energy
for the lowest rotation-aligned state in $^{190}$Hg
than in any other even-even Hg isotope
where superdeformation has been observed.

	In conclusion, the existence of an excited
SD band in $^{190}$Hg and its anomalous pattern
of decay to the lowest SD band have been confirmed.
 The energies of the transitions connecting
the two bands have been measured. From the
measured DCO ratios, the relative spins of
the two SD bands have been determined.  These
data provide further support for the hypothesis
that this excited SD band corresponds to rotations
built on an octupole-vibrational configuration.
 RPA calculations reproduce the experimental
data very well, and suggest that the octupole
phonon is rotationally aligned in the observed
range of spins.

	The authors wish to thank J. Kuehner and
G. Hackman for making available their software
for sorting data into a three-dimensional
histogram.
We would like to thank all those involved in the
setting up and commissioning
of Eurogam II,  especially Dominique Curien, Gilbert
Duchene and Gilles de France.
We would also like to thank the crew of the
Vivitron for the faultless
operation of the accelerator.
This work was supported by the
U.S. Department of Energy, Nuclear Physics
Division, under contract no. W-31-109-ENG-38.
The EUROGAM project is funded jointly by EPSRC
(UK) and IN2P3 (France). One of us (ANW)
acknowledges the receipt of an EPSRC
postgraduate studentship.


\begin{figure}
\caption{Sum of combinations of double coincidence
gates yielding events in which the excited SD band
was populated.  The pairs of coincidence gates
consisted of combinations of
transitions in the excited SD band with other
transitions in the excited band, plus transitions
in the excited band with transitions in the
lowest SD band.}
\label{Fig. 1}
\end{figure}

\begin{figure}
\caption{Proposed level-scheme of SD bands
in $^{190}$Hg.}
\label{Fig. 2}
\end{figure}

\begin{figure}
\caption{DCO ratios for selected transitions in $^{190}$Hg,
as defined in the text.  Circles, squares and triangles
are used, respectively, for E1 transitions between normally
deformed states, E2 transitions between states in the
lowest SD band, and transitions between the two SD bands.}
\label{Fig. 3}
\end{figure}

\begin{figure}
\caption{Comparison of observed and calculated
properties of SD bands in $^{190}$Hg versus
rotational frequency, $\hbar$$\omega$:  (a)
dynamic moment of inertia $\Im$$^{(2)}$; (b)
difference, E$^{\prime}$--E$^{\prime}$(lowest SD),
between Routhians of excited bands and that of
the lowest SD band, where
E$^{\prime}$$\equiv$E--$\omega$$_x$I$_x$.
Both vibrational and nonvibrational states are calculated
in the RPA, but only the lowest four excited states of
both signatures are shown here, and these are
the vibrational states at low frequencies (see text
for details).  The theoretical $\Im$$^{(2)}$ values
were obtained by calculating the difference in
$\Im$$^{(2)}$ between the excited and lowest SD bands
and adding a Harris polynomial as a reference.
The coefficients of the Harris polynomial
were $\Im$$_{0}$=82.6 MeV$^{-1}$
and $\Im$$_{1}$=113.0 MeV$^{-3}$, fitted to the levels
of the lowest SD band up to the 728 keV transition.}
\label{Fig. 4}
\end{figure}

\begin{table}
\begin{tabular}{|c|c|c|c|c|}
  & branching ratio &  &  & \\
 energy of & of inter- &partial half- & B(E1) & B(M1) \\
 transition (keV) & band transition & life (fs) &
             (W.u.$\times$10$^{-3}$) & (W.u.) \\ \hline
 911 & $>$0.5 & $<$200 & $>$1.4 & $>$0.15 \\
 864 & 0.35(4) & 260(60) & 1.2(3) & 0.13(3) \\
 812 & 0.29(4) & 260(70) & 1.5(4) & 0.16(4) \\
 757 & $<$0.3 & $>$200 & $<$2.4 & $<$0.26
\end{tabular}
\caption{
Information obtained for the transitions linking the
excited and lowest SD bands in this experiment.
The branching ratio shown in the second column is
the ratio between the out-of-band intensity and
the total intensity deexciting a particular level,
i.e. I$_{\gamma}$(I+1 $\rightarrow$ I) /
[I$_{\gamma}$(I+1 $\rightarrow$ I-1) +
I$_{\gamma}$(I+1 $\rightarrow$ I)].
The partial half-lives and transition strengths
are inferred from the data as explained in the
text.}
\label{Table I}
\end{table}

\end{document}